\pdfoutput=1
\documentclass[twoside,reqno]{HERON}
\usepackage{epsfig,cite,colordvi}
\usepackage{graphicx}
\usepackage{url}
\usepackage{amssymb,amsmath,amscd,epsf}
\usepackage{times}
\usepackage{makeidx}
\usepackage[utf8]{inputenc}
\makeindex
\begin{document}

\title{Bohr Hamiltonian with Hulthén plus ring-shaped potential for triaxial nuclei with deformation-dependent mass term }

\begin{start}

\author{A. Adahchour}{}, \coauthor{S. Ait El Korchi}{},\coauthor{ A. El Batoul}{}, \coauthor{ A. Lahbass}{},  \coauthor{ M. Oulne}{}

\address{High Energy Physics and Astrophysics Laboratory, Faculty of Sciences Semlalia, Cadi Ayyad University, P. O. B. 2390, Marrakech 40000, Morocco
}{}

\begin{Abstract}
In this work, we solve the eigenvalues problem with the Bohr collective Hamiltonian for triaxial nuclei within Deformation-Dependent Mass formalism (DDM) using the Hulthén potential . We shall call the solution developed here Z(5)-HDDM. Analytical expressions for energy spectra are derived by means of a recent version of the Asymptotic Iteration Method. The calculated numerical results are compared with the experimental data, and the model Z(5)-H using the Hulthén potential without DDM formalism as well as theoretical predictions of Z(5)-DDDM model with Davidson potential.

\end{Abstract}
\end{start}

\section{Introduction}
Usually, in Bohr's Hamiltonian, the mass parameter is considered as a constant. However, there is a growing evidence that this approximation may be inadequate. Several comparisons with experimental data have recently pointed out that the mass tensor of the collective Hamiltonian cannot be considered as a constant and should be taken as a function of the collective coordinates. Based on these proofs, a Bohr Hamiltonian with a mass depending on the collective variable can be treated. It should be noticed that the concept of a non-constant mass has been used long ago in quantum physics. But, it has been introduced for the first time in nuclear physics by D. Bonatsos et al \cite{BO4}\\
Atomic nuclei exhibit phase transitions as a function of the number of
nucleons. These phase transitions are of quantum type. Several critical
points of symmetries namely E(5), X(5), X(3), Z(5), Z(4) have been introduced.
In this work, we will focus on the critical point of symmetry Z(5) which
represents the transition from prolate axially symmetric SU(3) nuclei to oblate shapes. We will consider a Bohr Hamiltonian with the Hulthén potential including a mass parameter depending on the collective coordinate $\beta$. \\
The structure of the present work is as follows. In Section 2, the theoretical background of the elaborated model Z(5)-HDDM is briefly presented. In section 3, we give the obtained analytical expressions for the energy levels by means of Asymptotic Iteration Method. Section 4 contains results and discussion about the effect of DDM on the energy spectra of Xe and Pt isotopes.

\section{The Z(5)-HDDM model}
in the framework of Z(5), the original Bohr Hamiltonian \cite{B1} is
\begin{equation}
\begin{split}
		H_{B}=-\dfrac{\hbar^{2}}{2B} \bigg[\frac{1}{\beta^{4}}\frac{\partial}{\partial\beta}\beta^{4}\frac{\partial}{\partial\beta}+\frac{1}{\beta^{2}sin3\gamma}\frac{\partial}{\partial\gamma}\sin3\gamma\frac{\partial}{\partial\gamma}  \\ -\frac{1}{4\beta^{2}}\sum^{3}_{k=1}\frac{Q_{k}^{2}}{sin^{2}(\gamma-\frac{2\pi}{3}k)} \bigg]+ V(\beta,\gamma)
\end{split}
\label{HB1}
\end{equation}
Where B is the mass parameter, which is usually considered constant, $\beta$ and $\gamma$ are the usual collective coordinates ($\beta$ being a deformation coordinate measuring departure from spherical shape, and $\gamma$ being an angle measuring departure from axial symmetry), while $Q_{k}$ (k = 1, 2, 3) are the components of angular momentum in the intrinsic frame. 
\newline
Using a mass depending on the deformation coordinate $\beta$, 
\begin{equation}
   B(\beta)=\frac{B_{0}}{(f(\beta))^2}
\label{M1}
\end{equation}
where $B_{0}$ is the constant mass and $f(\beta)$ the deformation function. The Schrodinger equation corresponding to the Hamiltonian \eqref{HB1} is given by \cite{BO4}
\begin{equation}
\begin{split} 
H\Psi(\beta,\gamma,\theta_{i})=  \bigg[-\dfrac{1}{2}\dfrac{\sqrt f}{\beta^4}\dfrac{\partial}{\partial\beta}\beta^4 f\dfrac{\partial}{\partial\beta}\sqrt f    -\dfrac{f^2}{2\beta^2 sin3\gamma}\dfrac{\partial}{\partial\gamma}sin3\gamma \frac{\partial}{\partial\gamma}  \\
   +\; \frac{f^2}{8\beta^2}  \sum_{k=1,2,3}\frac {Q^{2}_{k}}{sin^2(\gamma-\frac{2}{3}\pi k)}+V_{eff} \bigg]\Psi(\beta,\gamma,\theta_{i}) =\varepsilon \Psi(\beta,\gamma,\theta_{i}) 
\end{split} 
\label{Ha1}                                                                                   
\end{equation}
Where $\theta_{i}$ are the Euler angles and the reduced energies $\varepsilon$, reduced potential $v(\beta,\gamma)$, effective potential $V_{eff}(\beta,\gamma)$ are respectively \\
\hspace*{2cm} $\varepsilon=\frac{B_{0}}{\hbar^2 }\;E$ , 
\hspace*{0.2cm} $v(\beta,\gamma)=\frac{B_{0}}{\hbar^2 }\;V(\beta,\gamma)$ \\
\hspace*{2cm} $V_{eff}(\beta,\gamma)=v(\beta,\gamma)+\frac{1}{4}(1-\delta-\lambda)f\nabla^2 f+\frac{1}{2}(\frac{1}{2}-\delta)(\frac{1}{2}-\lambda)(\nabla f)^2  $ \\
The function $f(\beta)$ depends only on the radial coordinate $\beta$, so only the $\beta$ part of the above equation is affected.

\section{Separable form of the Hamiltonian}
In order to achieve a separation of variables, we assume that the reduced potential $v(\beta,\gamma)$ depends on the variables $\beta$ and $\gamma$ and has the form \cite{Fo2,Fo3,Fo4,Bo3} \\
\begin{equation}
\hspace*{2cm}  v(\beta,\gamma)=u(\beta)+\frac{f^2}{\beta^2} \; w(\gamma)
\label{Pot1}
\end{equation}
with $w(\gamma)$ having a deep minimum at $\gamma$=$\frac{\pi}{6}$ and the wave functions have the form 
\begin{equation}
\hspace*{2cm} \Psi(\beta,\gamma,\theta_{i})=\xi(\beta) \; \Phi(\gamma,\theta_{i})
\label{Fon1}
\end{equation}
The separation of variables gives
\begin{equation}
\begin{split} 
\bigg[-\dfrac{1}{2}\dfrac{\sqrt f}{\beta^4}\dfrac{\partial}{\partial\beta}\beta^4 f\dfrac{\partial}{\partial\beta}\sqrt f + \frac{f^2}{2\beta^2}\Lambda+ \frac{1}{4}(1-\delta-\lambda)f\nabla^2 f \\   +\;\frac{1}{2}(\frac{1}{2}-\delta)(\frac{1}{2}-\lambda)(\nabla f)^2 +u(\beta) \bigg] \; \xi(\beta)=\varepsilon \;\xi(\beta)
\end{split} 
\label{Bb}                                                                                   
\end{equation}
and
\begin{equation}
\begin{split} 
\bigg[-\dfrac{1}{\sin3\gamma}\dfrac{\partial}{\partial\gamma}sin3\gamma \frac{\partial}{\partial\gamma}+ \frac{1}{4}  \sum_{k=1,2,3}\frac {Q^{2}_{k}}{sin^2(\gamma-\frac{2}{3}\pi k)} \\ + \; w(\gamma) \bigg]\Phi(\gamma,\theta_{i}) =\Lambda \; \Phi(\gamma,\theta_{i})
\end{split} 
\label{Bg}                                                                  
\end{equation}
where $\Lambda$ is the separation constant and equation \eqref{Bb} can be simplified by performing the derivations
\begin{equation}
\frac{1}{2}f^2\xi^{''}+\bigg(ff^{'}+\frac{2f^2}{\beta}\bigg)\xi^{'}+\bigg(\frac{(f^{'})^2}{8}+\frac{ff^{''}}{4}+\frac{ff^{'}}{\beta}-\frac{f^2}{2\beta^2}\Lambda+\varepsilon-v_{eff}\bigg)\xi=0
\label{Eqr1}
\end{equation}
with
\begin{equation}
v_{eff}=u(\beta)+\frac{1}{4}(1-\delta-\lambda)f(\frac{4f^{'}}{\beta}+f^{''})+\frac{1}{2}(\frac{1}{2}-\delta)(\frac{1}{2}-\lambda)(f^{'})^2
\end{equation}
In the present work, we use the Hulthén potential \cite{Hu1,Hu2} with a unit depth as in \cite{La1,Ma1}
\begin{equation}
u(\beta)=-\frac{1}{e^{\tau \beta}-1} 
\label{Hu}
\end{equation}
where $\tau=\frac{1}{b}$ is a screening parameter, and $b$ is the range of the potential. This potential has some properties, namely it behaves as a short-range potential for small values of $\beta$ and decreases exponentially for very large values of $\beta$. By inserting the function $R(\beta)=\beta^2\;\xi(\beta)$ in the radial equation \eqref{Eqr1}, one obtains
\begin{equation}
f^2 R^{''}+ 2ff^{'}R^{'}+\bigg(2\varepsilon-2(v_{eff}+\frac{f^2+\beta ff^{'}}{\beta^2}+\frac{f^2\Lambda}{2\beta^2}-\frac{(f^{'})^2}{8}-\frac{ff^{''}}{4})\bigg)R=0
\label{Eqr2}
\end{equation}
In order to make connection between our results and those obtained in ref. \cite{Ch2}, we have replaced $2\varepsilon$ by $\epsilon$ in the the above equation, so one obtains
\begin{equation}
f^2 R^{''}+ 2ff^{'}R^{'}+(\epsilon-2u_{eff})R=0
\label{Eqr3}
\end{equation}
where 
\begin{equation}
u_{eff}=v_{eff}+\frac{f^2+\beta ff^{'}}{\beta^2}+\frac{f^2\Lambda}{2\beta^2}-\frac{(f^{'})^2}{8}-\frac{ff^{''}}{4}
\label{Pef}
\end{equation}
The special form for the deformation function is
\begin{equation}
f(\beta)=1+a\beta^2   ,\;\;\;  a<<1
\label{Fdm}
\end{equation}
Using these forms for the potential and the deformation function in Eq. \eqref{Pef}, one obtains
\begin{equation}
2u_{eff}=k_{1}\beta^2+k_{0}+\frac{k_{-1}}{\beta^2}-\frac{1}{e^{\tau\beta}-1}
\end{equation}
Where 
\begin{subequations}
\begin{align}
k_{1}&=a^2 \bigg(5(1-\delta-\lambda)+(1-2\delta)(1-2\lambda)+4+\Lambda \bigg)  \\
k_{0}&=a \bigg(5(1-\delta-\lambda)+7+2\Lambda \bigg)  \\
k_{-1}&= 2+\Lambda
\end{align}
\label{kkk}
\end{subequations}
Equation \eqref{Eqr3} becomes
\begin{equation}
f^2 R^{''}(\beta)+ 2ff^{'}R^{'}(\beta)+\bigg(\epsilon-k_{1}\beta^2-k_{0}-\frac{k_{-1}}{\beta^2}+\frac{1}{e^{\tau\beta}-1} \bigg)R(\beta)=0
\label{Eqr4}
\end{equation}
To simplify equation \eqref{Eqr4}, we will proceed to a change of the function R($\beta$) by
\begin{equation}
R(\beta)=\frac{R(\beta)}{1+a\beta^2}
\end{equation}
So equation \eqref{Eqr4}, becomes 
\begin{equation}
\begin{split}
R^{''}(\beta)+\bigg(-\frac{k_{1}\beta^2}{(1+a\beta^2)^2}-\frac{2a}{1+a\beta^2}+\frac{\epsilon}{(1+a\beta^2)^2}+    \frac{1}{(1+a\beta^2)^2(e^{\tau\beta}-1)} \\ -\frac{k_{0}}{(1+a\beta^2)}-\frac{k_{-1}}{(1+a\beta^2)\beta^2}\bigg)R(\beta) =0
\end{split}
\label{Eqr5}
\end{equation}
From this equation, if we set the deformation parameter $a = 0$, we recover the equation (7) of ref. \cite{Ch2}. Because of the centrifugal potential and the form of the Hulthén one, the Schrodinger equation \eqref{Eqr5} cannot be solved analytically. So we will proceed to a rigorous approximation that allows to tackle this problem. For a small $\beta$ deformation, the centrifugal potential could be approximated by the following expression, as in refs. \cite{Ji1,Do1,So1}
\begin{equation}
\frac{1}{\beta^2}  \approx  \tau^2\frac{e^{-\tau\beta}}{(e^{-\tau\beta}-1)^2}
\end{equation}
This approximation is also valid for small values of the
screening parameter $\tau$. By using the new variable $y=e^{-\tau\beta}$, we obtain
\begin{equation}
\frac{1}{e^{\tau\beta}-1}=\frac{y}{1-y}\;\; , \;\;\;\; \beta=\frac{1-y}{\sqrt y \;\tau} \;\;,\;\;\; 1+a\beta^2=\frac{a(1-y)^2+y\tau^2}{y\tau^2}
\end{equation}
Rewriting equation \eqref{Eqr5} by using the new variable $y$, we obtain
\begin{equation}
\begin{split}
R^{''}(y)+\frac{1}{y}R^{'}(y)+\bigg(-\frac{(\epsilon-k_{0})\tau^2+(2-y)k_{1}}{(ay^2+(\tau^2-2a)y+a)^2}-\frac{k_{1}+2a}{y(ay^2+(\tau^2-2a)y+a)}  \\  +\frac{\tau^2\;y}{(1-y)(ay^2+(\tau^2-2a)y+a)^2}+\frac{\tau^4k_{-1}\;y}{(1-y)^2(ay^2+(\tau^2-2a)y+a)^2}\bigg)R(\beta)=0
\end{split}
\label{Eqr6}
\end{equation}
If $a = 0$, the dependence of the mass on the deformation is canceled, then we easily check that we get the equation (9) of ref. \cite{Ch2}. \\
The Schrodinger equation \eqref{Eqr6} cannot yet be solved analytically because of some terms. So, in the absence of a rigorous solution to this equation,
we can use a further approximation. For a small deformation parameter $a$ ($a<<1$), as a first approximation, we can neglected the following terms: $ay^2-2ay+a$, $k_{1}+2a$ and $\frac {k_{1}}{\tau^4y}$. So equation \eqref{Eqr6} becomes

\begin{equation}
\begin{split}
R^{''}(y)+\frac{1}{y}R^{'}(y)+\left[\frac{(\epsilon-k_{0})\tau^2+2k_{1}}{\tau^4y^2}-\frac{k_{-1}}{ y(1-y)^2}
+\frac{1}{\tau^2y(1-y)}\right]R(y)=0
\end{split}
\label{Eqr7}
\end{equation}
In order to transform the above differential equation to a more compact one, we use the following variables
\begin{equation}
\mu^2=-\frac{(\epsilon-k_{0})\tau^2+2k_{1}}{\tau^4} , \;\;\;\; \nu=\frac{1}{2}(1+\sqrt{1+4k_{-1}})
\end{equation}
So, the differential equation \eqref{Eqr7} becomes
\begin{equation}
\begin{split}
R^{''}(y)+\frac{1}{y}R^{'}(y)-\left[\frac{\mu^2}{y^2}+\frac{\nu^2-\nu}{ y(1-y)^2}
-\frac{1}{\tau^2y(1-y)}\right]R(y)=0
\end{split}
\label{Eqr8}
\end{equation}
To apply the asymptotic iteration method of refs. \cite{Ci1,Ci2}, the reasonable physical wave function that we propose is as follows
\begin{equation}
R(y)=y^{\mu}(1-y)^{\nu}\chi(y)
\end{equation}
For this form of the radial wave function, eq. \eqref{Eqr8} reads
\begin{equation}
\chi^{''}(y)=-\frac{\omega(y)}{\sigma(y)}\;\;\chi^{'}(y)-\frac{\kappa_{n}}{\sigma(y)}\;\;\chi(y)
\label{Echi}
\end{equation}
with
\begin{subequations}
\begin{align}
\omega(y)&=(2\mu+1)-(2\mu+2\nu+1)y  \\
\sigma(y)&=y(1-y) \\
\kappa_{n}&=\frac{1}{\tau^2}-\nu(2\mu+\nu)   
\end{align}
\end{subequations}
Equation \eqref{Echi} leads us directly to the energy eigenvalues  using the new generalized formula \cite{Boz1} which replaced the iterative calculations in the original AIM formulation \cite{Ci3}.
\begin{equation}
\kappa_{n}=-n\;\omega^{'}(y)-\frac{n(n-1)}{2}\;\;\sigma^{''}(y)
\label{Ekappa}
\end{equation}
The above formulation gives the energy spectrum of the $\beta$ equation
\begin{equation}
\begin{split}
\epsilon_{n}=-\left(\frac{\tau^2(n+\frac{1}{2}+\sqrt{\frac{1}{4}+k_{-1}})^2-1}{2\tau(n+\frac{1}{2}+\sqrt{\frac{1}{4}+k_{-1}})}     \right)^2 - 2\;\frac{k_{1}}{\tau^2}+k_{0}
\end{split}
\label{En}
\end{equation}
where $n$ is the principal quantum number and $k_{-1}$ is defined previously as a function of $\Lambda$, which represents the eigenvalues of the $\gamma$-vibrational plus rotational part of the Hamiltonian for triaxial nuclei. If we set the deformation parameter $a=0$, our energy spectrum formula eq. \eqref{En} is in agreement with the energy formula obtained in previous works refs. \cite{Ch2,Ik1,Ba2,Ag1}.\\
\\
For eq.\eqref{Bg}, which represents the $\gamma$ variable, we use a new generalized potential proposed in \cite{Ch3} that is inspired by a ring-shaped potential
\begin{equation}
w(\gamma)=\frac{c+s\; cos^{2}(3\gamma)}{sin^{2}(3\gamma)}
\label{Potg}
\end{equation}
where $c$ and $s$ are free parameters. Inserting this form of the potential in equation \eqref{Bg}, we get
\begin{equation}
\begin{split} 
\bigg[-\dfrac{1}{\sin3\gamma}\dfrac{\partial}{\partial\gamma}sin3\gamma \frac{\partial}{\partial\gamma}+ \frac{1}{4}  \sum_{k=1,2,3}\frac {Q^{2}_{k}}{sin^2(\gamma-\frac{2}{3}\pi k)} \\ +\; \frac{c+s\; cos^{2}(3\gamma)}{sin^{2}(3\gamma)}\bigg]\Phi(\gamma,\theta_{i}) =\Lambda \; \Phi(\gamma,\theta_{i})
\end{split} 
\label{Bg1}                                                                  
\end{equation}
Since the potential is minimal at $\gamma=\frac{\pi}{6}$, then the angular momentum term can be written as the form \cite{Fo1,Bo1}
\begin{equation}
\frac{1}{4}  \sum_{k=1,2,3}\frac {Q^{2}_{k}}{sin^2(\gamma-\frac{2}{3}\pi k)}\approx {\bf Q}^{2} -\frac{3}{4} Q_{1}^{2}
\label{Mo}                                                                  
\end{equation}
With ${\bf Q}^{2} = Q_{1}^{2}+Q_{2}^{2}+Q_{3}^{2}$ \\
One then take wave functions of the form 
\begin{equation}
\Phi(\gamma,\theta_{i})=\Gamma(\gamma) \; D_{M,\alpha}^{L}(\theta_{i})
\label{F1}                                                                  
\end{equation}
Thus, the separation of variables leads to the following set of differential equations
\begin{equation}
\bigg[-\dfrac{1}{\sin3\gamma}\dfrac{\partial}{\partial\gamma}sin3\gamma \frac{\partial}{\partial\gamma}+ \frac{c+s\; cos^{2}(3\gamma)}{sin^{2}(3\gamma)}\bigg]\; \Gamma(\gamma)=\Lambda^{'} \;\Gamma(\gamma)
\label{D1}                                                                  
\end{equation}

\begin{equation}
[Q^{2}-\frac{3}{4} Q_{1}^{2}]\; D_{M,\alpha}^{L}(\theta_{i})=\bar{\Lambda}\; D_{M,\alpha}^{L}(\theta_{i})
\label{D2}                                                                  
\end{equation}
where $D(\theta_{i})$ denotes Wigner functions of the Euler angles $\theta_{i}(i = 1, 2, 3)$, $L$ is the total angular momentum quantum number, while $M$ and $\alpha$ are the quantum numbers of the projections of angular momentum on the laboratory fixed $z$-axis and the body-fixed $x^{'}$-axis, respectively.\\ \\
Since the deformation function $f$ depends only on the radial coordinate $\beta$, only the $\beta$ part of the resulting equation was affected, the solution of the angular equation (see ref. \cite{Ch2} for details ) gives \\
\begin{equation}
\Lambda^{'} = 9n_{\gamma}(n_{\gamma}+1)+3\sqrt{c+s}(2n_{\gamma}+1)+c
\label{}                                                                  
\end{equation}
\begin{equation}
\bar{\Lambda} = \frac{L(L+4)+3n_{w}(2L-n_{w})}{4}
\label{La2}                                                                  
\end{equation}
where $n_{\gamma}$ is the quantum number related to $\gamma$-excitation, and $n_{w}$ the wobbling quantum number.  \\
Finally, the analytical expression of  $\Lambda$, which represents the eigenvalues of the $\gamma$-vibrational plus rotational part of the Hamiltonian for triaxial nuclei is
\begin{equation}
\Lambda = 9n_{\gamma}(n_{\gamma}+1)+3\sqrt{c+s}(2n_{\gamma}+1)+c +  \frac{L(L+4)+3n_{w}(2L-n_{w})}{4}
\label{}                                                                  
\end{equation}

\section{Numerical results}
The model $Z(5)-HDDM$ is applied for calculating the energies of the collective states  for the $^{126,128,130,132,134}Xe$ and $^{192,194,196}Pt$ isotopes. All these nuclei show the signature of the triaxial rigid rotor \cite{Ra,Da}
\begin{equation}
\Delta E=|E_{{2}^{+}_{g}}+E_{{2}^{+}_{\gamma}}-E_{{3}^{+}_{\gamma}}|=0
\label{De1}                                                                  
\end{equation}
This equation is used in an approximate way, because the experimental data for the eight nuclei lead to the values
\begin{equation}
\Delta E(KeV)=49, 17, 26, 162, 379, 8, 28, 29,
\label{De2}                                                                  
\end{equation}
for $^{126,128,130,132,134}Xe$ and $^{192,194,196}Pt$ isotopes, respectively. Referring to the values of equation \eqref{De2}, the $^{128,130}Xe$ and $^{192,194,196}Pt$ isotopes are good candidates for the triaxial rigid rotor model. Note that the formula \eqref{De1} serves here as a guide in choosing the candidate nuclei and therefore, we have also added the $^{126,132,134}Xe$ isotopes in our analysis. \\
The allowed bands (i.e. ground state, $\beta$ and $\gamma$) are labelled by the quantum numbers, $n$, $n_{w}$, $n_{\gamma}$ and $L$. As described in the framework of the rotation-vibration model \cite{Gr}, the lowest bands for $Z(5)$ are as follows 
\begin{enumerate}
\item The ground state band (gsb) is characterized by $n = 0$, $n_{\gamma}=0$, $n_{w}=0$
\item The $\beta$ band is characterized by $n = 1$, $n_{\gamma}=0$, $n_{w}=0$.
\item The $\gamma$ band composed by the even $L$ levels with $n =0$, $n_{\gamma}=0$, $n_{w}=2$ and the odd $L$ levels with $n = 0$, $n_{\gamma}=0$, $n_{w}=1$.
\end{enumerate}
The energy spectrum is given by equation \eqref{En} and depending on four parameters, namely the screening parameter $\tau$ in the $\beta$ potential,  the ring-shape parameters $c$ and $s$ of the $\gamma$ potential and the deformation parameter $a$. Our task is to fit these parameters to reproduce the experimental data by applying a least-squares fitting procedure for each considered isotope. We evaluate the root mean square (rms) deviation between the theoretical values and the experimental data by
\begin{equation}
\sigma=\sqrt{\frac{\sum_{i=1}^{m}(E_{i}(exp)-E_{i}(th))^2}{(m-1)E(2_{1}^{+})^2}}
\label{De3}                                                                  
\end{equation}
where $E_{i}(exp)$ and $E_{i}(th)$ represent the theoretical and experimental energies of the $\textit i^{th}$ level, respectively, while $m$ denotes the number of states. $E(2_{1}^{+})$ is the energy of the first excited level of the ground state band.
The corresponding free parameters ($\tau$, $c$, $s$) and the deformation parameter $a$ are listed in table \ref{table:T1}. 
In this table, we give the fitted parameters allowing to reproduce the experimental data \cite{ww1} and $Z(5)$ model \cite{Bo2}. The results presented here have been obtained for $\delta=\lambda=0$. Different choices for $\delta$ and $\lambda$   lead to a renormalization of the parameter values $\tau$, $c$, $s$ and $a$ , so the predicted energy levels remain exactly the same. In table \ref{table:T2}, we compare the quality measure $\sigma$ of our results \textit{Z(5)-HDDM} with \textit{Z(5)-H} \cite{Ch2}, $Z(5)$ \cite{Bo2} and $Z(5)-DDDM$ \cite{BO4}.\\
Let's just point out that with Davidson's potential \cite{BO4}, for $^{192}Pt$ isotope, the highest level of beta band is 0 (4 in our case), for $^{194}Pt$  isotope, the highest level of beta band is 5 (8 in our case) and for $^{196}Pt$  isotope, the highest levels of even gamma band is 2 and odd gamma band is 6 (respectively 4 and 8 in our case).

\begin{table}[h]
 	
 	\small\noindent\tabcolsep=10pt
 	 
 	\begin{tabular}{ c c c c c c c c c}
 		\hline 
 		
 		\hline
 		\\[-8pt]
 		nuclei \qquad&${\tau}$& $c$& $s$& $a$& $L_{g}$& $L_{\beta}$& $L_{\gamma}$& $m$ \\
 		\hline
 		\\[-8pt]
 		{$^{126}Xe$} \quad&0.071& 8& 192& 0.0025& 12& 4& 9& 16 \\
 		{$^{128}Xe$} \quad&0.050& 2& 140& 0.0000& 10& 2& 7& 12 \\
 		{$^{130}Xe$} \quad&0.010& 0& 140& 0.0000& 14& 0& 5& 11 \\
 		{$^{132}Xe$} \quad&0.080& 72& 226& 0.0000& 6&  0& 5& 7 \\
 		{$^{134}Xe$} \quad&0.080& 78& 187& 0.0000& 6&  0& 5& 7 \\
 		{$^{192}Pt$} \quad&0.050& 19& 73& 0.0010& 10& 4& 8& 14 \\
 		{$^{194}Pt$} \quad&0.059& 6& 195& 0.0030& 10& 4& 8& 13 \\
 		{$^{196}Pt$} \quad&0.086& 7& 120& 0.0059& 10& 4& 8& 13 \\
 		{$Z(5)$}     \quad&0.039& 11& 406& -& 14& 4& 9& 17 \\
 		\hline
 	\end{tabular}
 	\caption{Free and deformation parameters values fitted to the experimental data \cite{ww1} and $Z(5)$ model \cite{Bo2}. $L_{g}$, $L_{\beta}$ and $L_{\gamma}$ characterize the angular momenta of the highest levels of the ground state, $\beta$ and $\gamma$ bands respectively, included in the fit, while $m$ the total number of experimental states involved in the rms fit.}
\label{table:T1}
 \end{table}

\begin{table}[h]	
 	\small\noindent\tabcolsep=13pt
 	\begin{tabular}{ c c c c c}
 		\hline 
 		
 		\hline
 		\\[-8pt]
 		nuclei \qquad&${Z(5)-HDDM}$&${Z(5)-H}$&${Z(5)}$&$Z(5)-DDDM$  \\
 		\hline
 		\\[-8pt]
 		{$^{126}Xe$} \quad&0.716& 0.835& 1.082& 0.584 \\
 		{$^{128}Xe$} \quad&0.508& 0.508& 0.802& 0.431 \\
 		{$^{130}Xe$} \quad&0.443& 0.443& 1.564& 0.347 \\
 		{$^{132}Xe$} \quad&0.181& 0.181& 1.013& 0.467 \\
 		{$^{134}Xe$} \quad&0.123& 0.123& 1.524& 0.685 \\
 		{$^{192}Pt$} \quad&0.517& 0.521& 0.886& 0.681 \\
 		{$^{194}Pt$} \quad&0.544& 0.553& 0.973& 0.667 \\
 		{$^{196}Pt$} \quad&0.602& 0.718& 1.448& 0.639 \\
 		\hline
 	\end{tabular}
 	\caption{The root mean square (rms) deviation  between experimental data \cite{ww1} and the theoretical results corresponding to $Z(5)-HDDM$ model, $Z(5)-H$ model \cite{Ch2} and $Z(5)-DDDM$ model \cite{BO4} of given isotopes.}
\label{table:T2}
 \end{table}

\section*{Conclusion}
In this work, we have solved the eigenvalues problem with the Bohr collective Hamiltonian for triaxial nuclei within deformation-dependent mass formalism. 
Using the potential of Hulth\'en, we have improved the accuracy of the results obtained in reference \cite{Ch2} except for isotopes with the deformation parameter $a$ equal to $0$. From the comparison with the results obtained in Ref\cite{BO4} using the potential of Davidson, one can conclude that the results for $^{126,128,130}Xe$  isotopes are less accurate but for $^{132,134}Xe$ and $^{192,194,196}Pt$ isotopes, we have an improvement in accuracy. Our results confirm those obtained in Ref \cite{BO4} concerning the vibratory nature of the isotopes $^{128,130,132,134}Xe$.  \\
It has been shown that, whether with Hulth\'en's potential or with Davidson's potential, the results have been improved in general, compared whith the infinite square well.  Consequently, for a better description of the experimental data or to enlarge the palette of applications, it is important to test other more flexible potentials. 
\section*{Acknowledgements}
A. Adahchour acknowledges the financial support(Type A) of Cadi Ayyad University (Morocco). Also, he would like to thank and warmly congratulate all the members of the organizing committee for their welcome and their hospitality.


\begin{thebibliography}{99}
  \bibitem{BO4} D. Bonatsos, P. E. Georgoudis, D. Lenis, N. Minkov and C. Quesne, Phys. Rev. C83, 044321, (2011).
  \bibitem{B1} A. Bohr, Mat. Fys. Medd. K. Dan. Vidensk. Selsk. 26, no. 14 (1952).
  \bibitem{Fo2} L. Fortunato, Eur. Phys. J. A26 (s01), 1 (2005).
	\bibitem{Fo3} L. Fortunato, Phys. Rev. C70, 011302 (2004).
	\bibitem{Fo4}  L. Fortunato, S. De Baerdemacker, and K. Heyde, Phys. Rev. C74, 014310 (2006).
	\bibitem{Bo3} D. Bonatsos, E. A. McCutchan, N. Minkov, R. F. Casten, P. Yotov, D. Lenis, D. Petrellis, I. Yigitoglu, Phys. Rev. C76, 064312 (2007).
	\bibitem{Hu1} L. Hulthen, Ark. Mat. Astron. Fys. A 28, 5 (1942).	
	\bibitem{Hu2} L. Hulthen, Ark. Mat. Astron. Fys. B29, 1 (1942).
	\bibitem{La1} U. Laha, C. Bhattacharyya, K. Roy, B. Talukdar, Phys. Rev. C38, 558 (1988).
	\bibitem{Ma1} P. Matthys, H. De Meyer, Phys. Rev. A38, 1168 (1988).
  \bibitem{Ch2} M. Chabab, A. Lahbas, and M. Oulne, Eur. Phys. J. A (2015) 51: 131.
  \bibitem{Ji1} C.S. Jia, T. Chen, L.G. Cui, Phys. Lett. A373, 1621 (2009).
	\bibitem{Do1} S.H. Dong, W.C. Qiang, G.H. Sun, V.B. Bezerra, J. Phys.A: Math. Theor. 40, 10535 (2007).
	\bibitem{So1} A. Soylu, O. Bayrak, I. Boztosun, J. Phys. A: Math. Theor. 41, 065308 (2008).	
	\bibitem{Ci1}  H. Ciftci, R.L. Hall, N. Saad, J. Phys. A36, 11807 (2003).
	\bibitem{Ci2} H. Ciftci, R.L. Hall, N. Saad, J. Phys. Math. Gen. A38, 1147 (2005).
	\bibitem{Boz1} I. Boztosun, M. Karakoc, Chin. Phys. Lett. 24, 3028 (2007).
	\bibitem{Ci3} H. Ciftci, R.L. Hall, N. Saad, J. Phys. A36, 11807 (2003).
	\bibitem{Ik1} S.M. Ikhdair, R. Sever, J. Math. Chem. 42, 461 (2007).
	\bibitem{Ba2} O. Bayrak, G. Kocak, I. Boztosun, J. Phys. A: Math. Gen.39, 11521 (2006).
	\bibitem{Ag1} D. Agboola, Commun. Theor. Phys. 55, 972 (2011).
	\bibitem{Ch3} M. Chabab, A. Lahbas, M. Oulne, Int. J. Mod. Phys. E21, 10 (2012).
	\bibitem{Fo1} L. Fortunato, Phys. Rev. C70, 011302 (2004).
	\bibitem{Bo1} D. Bonatsos, D. Lenis, N. Minkov, D. Petrellis, P.P. Raychev, P.A. Terziev, Phys. Lett. B584, 40 (2004).
	\bibitem{Me} J. Meyer-ter-Vehn, Nucl. Phys. A249, 111 (1975).
	\bibitem{B2} A. Bohr, B.R. Mottelson, Nuclear Structure Vol. II: Nuclear Deformations (Benjamin, New York, 1975).
	\bibitem{Ra} A.A. Raduta, P. Buganu, Phys. Rev. C83, 034313 (2011).
	\bibitem{Da} A.S. Davydov, G.F. Fillipov, Nucl. Phys. 8,237 (1958).
	\bibitem{Gr} W. Greiner, J.A. Maruhn, Nuclear Models (Springer, Berlin, 1996).
	\bibitem{ww1} http://www.nndc.bnl.gov/nndc/ensdf/.
	\bibitem{Bo2} D. Bonatsos, D. Lenis, D. Petrellis, P.A. Terziev, Phys. Lett. B588, 172 (2004).
	
		
\end{thebibliography}
\end{document}